\documentstyle{mn}

\newif\ifAMStwofonts

\ifoldfss

  \ifCUPmtlplainloaded \else
    \NewTextAlphabet{textbfit} {cmbxti10} {}
    \NewTextAlphabet{textbfss} {cmssbx10} {}
    \NewMathAlphabet{mathbfit} {cmbxti10} {} 
    \NewMathAlphabet{mathbfss} {cmssbx10} {} 
  \fi
  \ifAMStwofonts
    \ifCUPmtlplainloaded \else
      \NewSymbolFont{upmath} {eurm10}
      \NewSymbolFont{AMSa} {msam10}
      \NewMathSymbol{\upi}     {0}{upmath}{19}
      \NewMathSymbol{\umu}     {0}{upmath}{16}
      \NewMathSymbol{\upartial}{0}{upmath}{40}
      \NewMathSymbol{\leqslant}{3}{AMSa}{36}
      \NewMathSymbol{\geqslant}{3}{AMSa}{3E}

    \fi
  \fi
\fi 

\ifnfssone
  \newmathalphabet{\mathit}
  \addtoversion{normal}{\mathit}{cmr}{m}{it}
  \addtoversion{bold}{\mathit}{cmr}{bx}{it}

  \newmathalphabet{\mathbfit} 
  \addtoversion{normal}{\mathbfit}{cmr}{bx}{it}
  \addtoversion{bold}{\mathbfit}{cmr}{bx}{it}
  \newmathalphabet{\mathbfss} 
  \addtoversion{normal}{\mathbfss}{cmss}{bx}{n}
  \addtoversion{bold}{\mathbfss}{cmss}{bx}{n}
  \ifAMStwofonts
    \ifCUPmtlplainloaded \else
      \UseAMStwoboldmath
      \makeatletter
      \new@mathgroup\upmath@group
      \define@mathgroup\mv@normal\upmath@group{eur}{m}{n}
      \define@mathgroup\mv@bold\upmath@group{eur}{b}{n}
      \edef\UPM{\hexnumber\upmath@group}
      \new@mathgroup\amsa@group
      \define@mathgroup\mv@normal\amsa@group{msa}{m}{n}
      \define@mathgroup\mv@bold\amsa@group{msa}{m}{n}
      \edef\AMSa{\hexnumber\amsa@group}
      \makeatother
      \mathchardef\upi="0\UPM19
      \mathchardef\umu="0\UPM16
      \mathchardef\upartial="0\UPM40
      \mathchardef\leqslant="3\AMSa36
      \mathchardef\geqslant="3\AMSa3E

    \fi
  \fi
\fi 

\ifnfsstwo

  \DeclareMathAlphabet{\mathbfit}{OT1}{cmr}{bx}{it}
  \SetMathAlphabet\mathbfit{bold}{OT1}{cmr}{bx}{it}
  \DeclareMathAlphabet{\mathbfss}{OT1}{cmss}{bx}{n}
  \SetMathAlphabet\mathbfss{bold}{OT1}{cmss}{bx}{n}
  \ifAMStwofonts
    \ifCUPmtlplainloaded \else
      \DeclareSymbolFont{UPM}{U}{eur}{m}{n}
      \SetSymbolFont{UPM}{bold}{U}{eur}{b}{n}
      \DeclareSymbolFont{AMSa}{U}{msa}{m}{n}
      \DeclareMathSymbol{\upi}{0}{UPM}{"19}
      \DeclareMathSymbol{\umu}{0}{UPM}{"16}
      \DeclareMathSymbol{\upartial}{0}{UPM}{"40}
      \DeclareMathSymbol{\leqslant}{3}{AMSa}{"36}
      \DeclareMathSymbol{\geqslant}{3}{AMSa}{"3E}

    \fi
  \fi
\fi 

\ifCUPmtlplainloaded \else
  \ifAMStwofonts \else 
    \def\upi{\pi}
    \def\umu{\mu}
    \def\upartial{\partial}
  \fi
\fi

\pagerange{}    
\pubyear{} \volume{}

\title[Spectral variability in discs]{Spectral variability in transonic
discs around black holes}

\author[L. Zampieri, R. Turolla, E. Szuszkiewicz]{Luca
Zampieri$^1$,
Roberto Turolla$^1$ and Ewa Szuszkiewicz$^{1,2,3,4}$\\
$^1$Department of Physics, University of
Padova, via Marzolo 8, 35131 Padova, Italy \\
$^2$Institute of Physics, University of Szczecin, ul. Wielkopolska 15,
70-451 Szczecin, Poland \\
$^3$Torun Centre for Astronomy, Nicolaus Copernicus University,
ul. Gagarina 11, 87-100 Toru\'n, Poland \\
$^4$International School for Advanced Studies, via Beirut 2-4,
34013 Trieste, Italy}

\pagerange{\pageref{firstpage}--\pageref{lastpage}}

\pubyear{2001}

\begin{document}

\maketitle

\label{firstpage}

\begin{abstract}
Transonic discs with accretion rates relevant to intrinsically
bright Galactic X-ray sources ($L\approx 10^{38}$--$10^{39} \ {\rm
erg\,s}^{-1}$) exhibit a time dependent cyclic behaviour due to
the onset of a thermal instability driven by radiation pressure.
In this paper we calculate radiation spectra emitted from
thermally-unstable discs to provide detailed theoretical
predictions for observationally relevant quantities. The emergent
spectrum has been obtained by solving self-consistently the
vertical structure and radiative transfer in the disc atmosphere.
We focus on four particular stages of the disc evolution,
the maximal evacuation stage and three
intermediate stages during the replenishment phase. The disc is
found to undergo rather dramatic spectral changes during the
evolution, emitting mainly in the 1--10 keV band during outburst
and in the 0.1--1 keV band off-outburst. Local spectra, although
different in shape from a blackbody at the disc effective
temperature, may be characterized in terms of a hardening factor
$f$. We have found that $f$ is rather constant both in radius and
in time, with a typical value $\sim 1.65$.
\end{abstract}


\section{Introduction}
\label{intro}

Observations of Galactic black hole candidates (BHCs) and active
galactic nuclei (AGNs) have gone a long way in obtaining very
detailed information on these sources. The analysis of X-ray data
revealed the existence of at least four distinct spectral states
in BHCs, defined on the basis of the spectral shape and the flux
level (see e.g. Tanaka \& Lewin \cite{tl95:1995}; Narayan,
Mahadevan \& Quataert \cite{namaqua99:1999} for reviews). In the
high (soft) state the spectrum is dominated by a soft thermal
component, commonly attributed to a standard accretion disc. In
the low (hard) state a power-law high energy tail, probably
originated by the comptonization of soft photons by a hotter
medium, dominates the spectrum. Usually, as the flux decreases,
the power-law component becomes more and more pronounced, while
the thermal one weakens and eventually disappears. Optical/UV/X-ray
AGN spectra
consist of three main components: a big blue bump, a soft X-ray
excess, a hard power-law tail with a reflection hump and a
high-energy cut-off (see e.g. Zdziarski et al. \cite{zdz96:1996}
for a review). As it was realized long ago, by shifting a typical
AGN spectrum to higher energies by a factor $\approx 100$, such
that the big blue bump falls in the 1-10 keV band, one obtains a
X-ray spectrum which closely resembles that of a typical BHC in
an intermediate state. The actual similarity between these two
classes of sources (modulo the energy shift) is corroborated by a
number of observational evidences (see e.g. Grebenev et al.
\cite{g93:1993};  Grebenev, Sunyaev \& Pavlinsky \cite{g97:1997}
for the properties of power-law tails and Zdziarski, Lubi\'nski
\& Smith \cite{zls:1999} for the spectral index-Compton
reflection hump correlation).

In the last few years, a great wealth of information on the
physical conditions in BHCs has come from the study of their
variability. Besides transitions among different spectral states
(on time-scales $\approx 1$ hour-1 month), time analysis revealed
much more rapid variations, ranging from very low (mHz) to
high-frequency (100 Hz) QPOs. In some sources (e.g. GRS 1915+105)
phase lags between photons in different X-ray bands have been
detected, with typical values in the range 10-1000 ms (Cui et al.
\cite{cui99:1999}; Reig et al. \cite{reig00:2000}; Cui, Zhang \&
Chen \cite{cui00:2000}; Pottschmidt et al. \cite{pott00:2000}).
If the characteristic times scale linearly with the black hole
mass, as one naively expects, the shortest variability in AGNs
should occur on $\sim$ hours-days. Up to now, no definite
evidence for QPOs nor phase lags has been reported in AGNs,
although multi-wavelength observations of some active galaxies
seem to indicate a correlated variability between the optical-UV
and X-ray fluxes (Clavel et al. \cite{clav92:1992}; Edelson et
al. \cite{edel96:1996}; Nandra et al. \cite{nan98:1998}). Time
delays of a few days between the optical-UV and X-ray bands could
be detectable, providing a powerful tool to probe the geometry of
the accretion flow in the nuclear regions, as recently suggested
by Zampieri et al. \cite{zampeta00:2000}.

The analogies between the spectral properties in the
optical/UV/X-ray bands and, possibly, the variability patterns in
BHCs and AGNs suggest that in both classes of sources the same
basic mechanism might be responsible for the observed emission.
Much effort was devoted in recent years to envisage a ``unified''
scenario that has been partially successful in explaining the
different spectral states in BHCs. The current picture is based
on a two-component accretion flow: a standard disc and a hotter
phase which could be advection-dominated (see e.g. Narayan,
Mahadevan \& Quataert \cite{namaqua99:1999}). The
different states are then produced by the varying importance of
the two components. However, specific models do not provide a
consistent description for the structure and evolution of the
accretion flow.

It has been known since a long time that the inner regions of a
standard disc are thermally and viscously unstable (e.g. Pringle, Rees
\& Pacholczyk \cite{PRP:1973}; Shapiro, Lightman \& Eardely
\cite{sle76:1976}; Ichimaru \cite{ichi:1977}). This suggested
the intriguing possibility that the observed spectral transitions may
be triggered by such instabilities. A definite answer should be sought
in terms of a time-dependent hydrodymical approach, such as that
adopted by Honma, Matsumoto \& Kato \cite{hmk:1991} and Szuszkiewicz
\& Miller \cite{sm97:1997}, \cite{sm98:1998}. In particular, the latter
authors have calculated the non-linear (global) evolution of transonic
accretion discs produced by the onset of a thermal instability
driven by radiation pressure and have shown that the evolution
follows a cyclic pattern, passing through successive stages of
evacuation and replenishment of the inner disc region (see also
Takeuchi \& Mineshige \cite{tami:1998}).

Ultimately, in order to get detailed theoretical predictions for
the observed quantities, it is necessary to compute the spectra
emitted during the instability cycle. Model spectra
should properly take into account for the effects of the disc
atmosphere, through which energy deposition is likely to take
place at different heights. Models of disc atmospheres have been
already presented (LaDous \cite{LD89:1989}; Shimura \& Takahara
\cite{st93:1993}) and some thought has been given to the problem
of energy deposition (Shaviv \& Wehrse \cite{sw86:1986}; Hubeny
\cite{H90:1990}).  The first thorough investigation of the vertical
structure and emission spectrum of a thin accretion disc has been
presented by Shimura \& Takahara \cite{st95:1995}, who found that, for
intermediate accretion rates, local spectra, emitted from each disc
annulus, can be approximated by a Planckian with an average hardening
factor $f \sim 1.7$. Investigations of the spectra
emitted by  steady-state, dissipative accretion discs in AGNs have 
been carried out
by Sincell \& Krolik \cite{sk:1998} (standard discs) and Wang et al.
\cite{wangetal:1999} (slim discs). Recently, Hubeny and co-workers
(\cite{H00:2000} and references therein) reported on the progress in
their long term programme to construct detailed spectra from
stationary, thin accretion discs in AGNs.

In this paper we compute spectra under the
slim-disc approximation, which is appropriate for modeling
intrinsically bright sources and allows us to follow the
non-stationary evolution of a disc subject to a thermal
instability. We will assume that energy deposition takes place
entirely in the disc. More complex situations in which energy
deposition occurs also in the atmosphere will be considered in a
follow-up paper. This work is part of an ongoing research
programme dedicated to the analysis of hydrodynamical
instabilities in slim accretion discs and to the modeling of the
spectral and variability properties of BHCs, in particular of the
microquasar GRS 1915+105. In this paper we compute the spectra
that emerge from slim discs during their instability cycle and
discuss their possible relevance for the soft spectral component
of GRS 1915+105. The paper is organized as follows. In Section
\ref{hydro} we describe the structure and evolution of thermally
unstable transonic disc, concentrating on those properties
relevant to spectral calculations. In Section \ref{radtra} our
approach to the solution of radiative transfer in the disc
atmosphere is outlined. Computed spectra are presented in
Section \ref{result}. Discussion and conclusions follow in Section
\ref{conclu}.

\section{Radial structure of a thermally unstable transonic accretion disc}
\label{hydro}

\begin{figure}
\vskip 7.8cm
\includegraphics{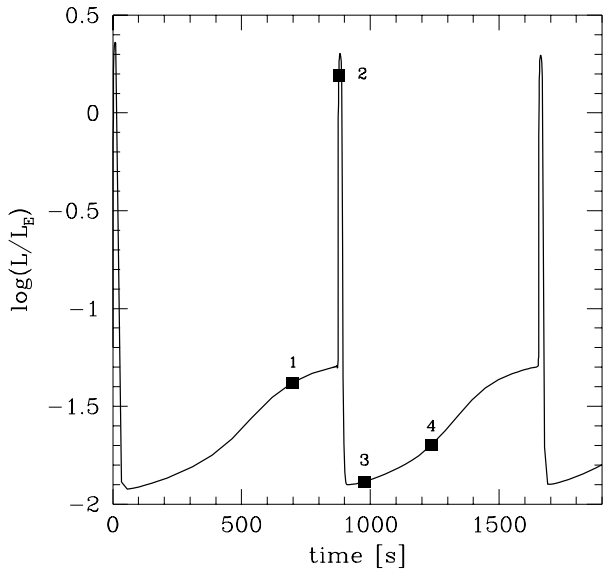}
\caption{\label{light}{The bolometric light curve with
the positions of four particular evolutionary stages described in the text.
}}
\end{figure}

Thermally unstable discs may undergo a limit-cycle behaviour with
successive evacuation and refilling of their central parts. By means
of a full time-dependent hydrodynamical simulation, Szuszkiewicz \&
Miller \cite{sm98:1998} explored in detail a model with black hole
mass $M =10 M_{\odot}$, initial accretion rate $\dot M= 0.96$ (in
Eddington units) and viscosity parameter $\alpha =0.1$. In this
particular model the cycle has a duration of about 780 s and can be
divided into two distinct parts: a very short ($\sim 20$ s) outburst
phase, during which the inner part of the disc is evacuated, and a
quiescent stage (corresponding to the slow refilling) lasting for
$\sim 760$ s. We have chosen four particular snapshots during the disc
evolution for which to calculate emergent spectra. They correspond to
the end of the refilling phase, just before the outburst starts (model
1), the peak of the outburst (the maximal evacuation stage, model 2),
the beginning (model 3), and half way through the refilling phase
(model 4). The theoretical bolometric light curve together with the positions
of the four chosen snapshots is shown in Fig. \ref{light}.
In this section we discuss the radial structure of the disc
for each of the four stages.

\begin{figure}
\vskip 7.8cm
\includegraphics{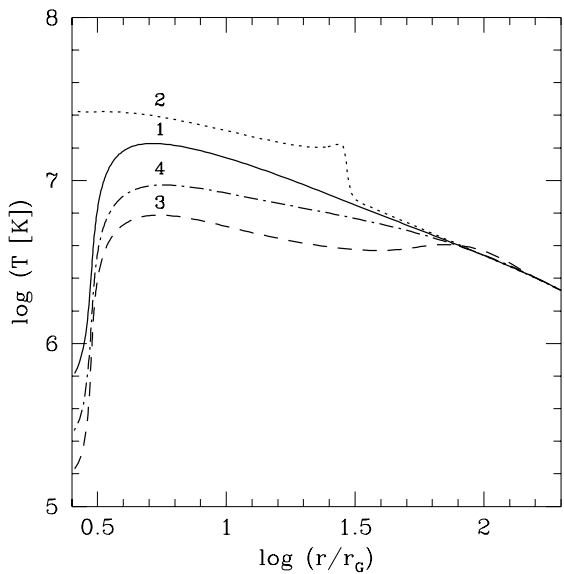}
\caption{\label{temp}{The radial profile of the midplane electron temperature for
the four particular evolutionary stages discussed in the text.
}}
\end{figure}

\begin{figure}
\vskip 7.5cm
\includegraphics{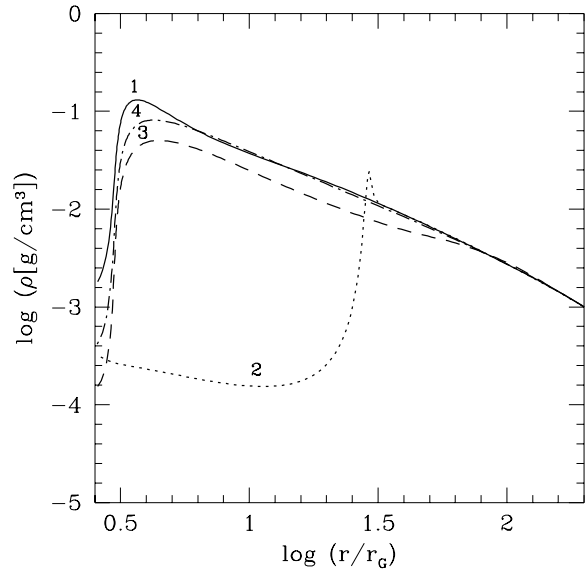}
\caption{\label{density}{
Same as in Fig. \ref{temp} for the volume density.
}}
\end{figure}

\begin{figure}
\vskip 7.5cm
\includegraphics{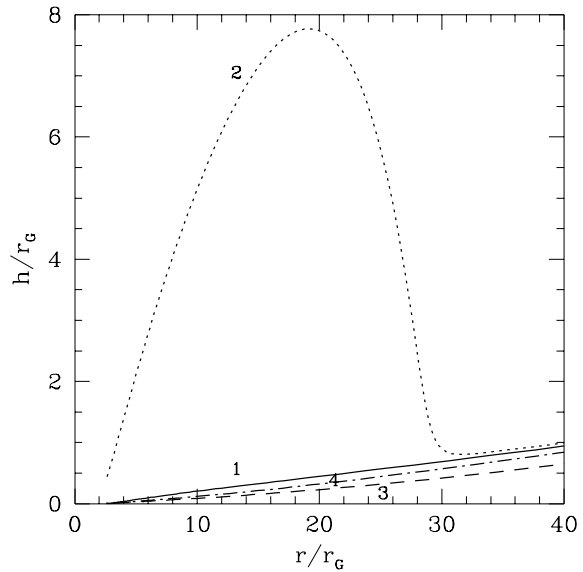}
\caption{\label{thick}{
Same as in Fig. \ref{temp} for the disc half-thickness.
}}
\end{figure}

The electron temperature distributions for models 1-4 are shown in
Fig. \ref{temp}; unless stated otherwise, hereafter all lengths
are in units of the Schwarzschild radius, $r_G = 2GM/c^2$. Before
the instability sets in (model 1) the temperature profile is that
given by the solid line, labeled 1 in Fig. \ref{temp}.  At the
onset of the instability, two density waves are sent out from the
locally-unstable region, one moving inwards and the other
outwards. At the same time the temperature rises significantly in
the unstable region.  As the outgoing wave progresses the
temperature peak is reduced (model 2) but remains still above
that of model 1. Another effect of the propagating wave is to push
a significant amount of matter outwards leaving behind an
underdense region. This can be seen in Fig. \ref{density} where
the volume density is plotted against radius.  The outgoing wave
heats the material through which it passes, causing the perturbed
part of the disc to swell up, as shown in Fig. \ref{thick}.
During the evacuation phase the disc temperature is high.
However, once the wavefront has moved beyond the linearly unstable
region, the disc cannot remain at such a high
temperature for a very long time. Eventually the front weakens and the temperature falls
dramatically (model 3). After this, the underdense region begins
to fill up on a viscous time-scale. This increase in density is
associated with a progressive rise in the temperature.
\phantom{}

\begin{figure}
\vskip 7.5cm
\includegraphics{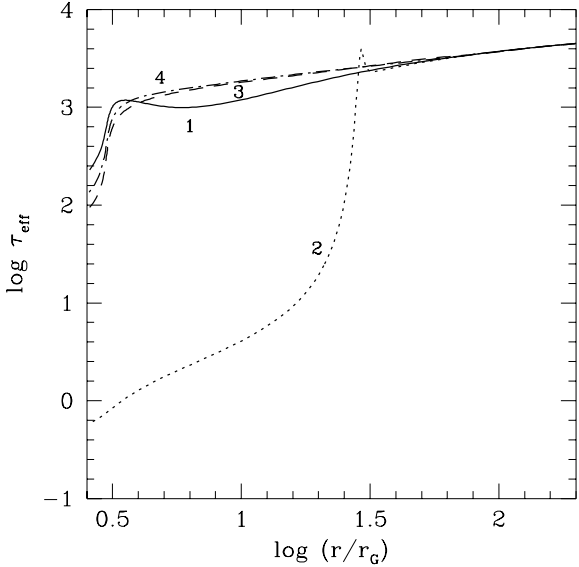}
\caption{\label{taueff}{Same as in Fig. \ref{temp} for the effective
optical depth.}
}
\end{figure}

\begin{figure}
\vskip 7.5cm
\includegraphics{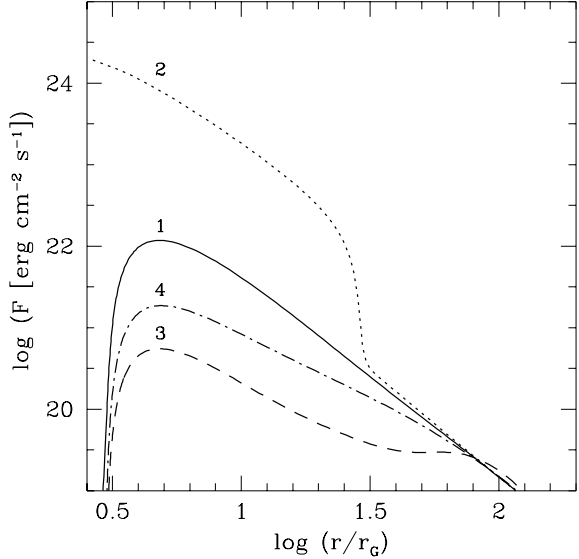}
\caption{\label{flux}{Same as in Fig. \ref{temp} for the emitted flux.}
}
\end{figure}

A quantity of immediate interest for spectral calculations is the
effective optical thickness of the disc,
$\tau_{eff} = [\tau_{a}(\tau_{es} + \tau_a)]^{1/2}$; here $\tau_{es}$ and
$\tau_a$ are the scattering and absorption depths
respectively. As shown in Fig \ref{taueff}, during the
the maximal evacuation stage (model 2) the effective optical depth
drops slightly below unity in the disc inner region.
However, $\tau_{eff}$ is sufficiently large to safely assume that the
radiation field in the disc is in LTE. Then, close to the equatorial
plane the radiation intensity approaches a blackbody.  In traversing
the disc atmospheric layers radiation goes out of equilibrium and
scattering effects are likely to change the spectrum, as we will
discuss in more detail in Section \ref{result}.  The total flux
distribution for each stage of the evolution is shown in
Fig. \ref{flux}. Neglecting relativistic effects, the total luminosity
of the disc in the four different stages $L_{disc}$ can be computed
integrating the flux over the disc surface (see Table
\ref{tabmod}). It should be noted that $L_{disc}$ becomes larger than the Eddington
limit in the outburst phase. Super-Eddington luminosities may indeed be
produced during the
non-steady evolution and/or in a non-spherical flow. In this particular case,
this happens because of the significant transient increase in the disc
temperature. The flux together with the disc thickness
will be used in the next section to calculate the emergent spectra.

\section{Radiative transfer in the disc atmosphere}
\label{radtra}

In this section we outline the numerical method used in computing
the evolution of the radiation spectrum emitted by the transonic
accretion disc. As discussed previously (see Section \ref{hydro}), we
consider four stages that correspond to
the maximal evacuation stage (model 2) and three
replenishment stages (model 1, 3 and 4).

The radial grid of each numerical disc model is made up of
$\sim$250 points dividing the integration domain,
$r_{in} = 2.5<r<r_{out} = 2.8\times 10^5$, into
a succession of comoving annular zones, each one
$11\%$ more massive than the one interior to it. The mass of the
innermost zone was determined in accordance with numerical convenience.
In order to compute the spectrum of these models, we introduce
coarser radial zones (or rings) whose width is fixed by the
requirement that their contribution to the total luminosity is $\sim
5\%$. We have found more convenient to keep the ring boundaries
coincident with the mesh points of the hydrodynamic calculation, avoiding
unnecessary interpolations. The luminosity $L_r$ and average height
$h_r$ of each ring are then computed as

\begin{eqnarray}
&& L_r = \sum_i 2\pi r_{i-1/2} \Delta r_i r_G^2 F_{i-1/2} \\
&& h_r = \sum_i h_i/N
\end{eqnarray}
where $r_i$ and $\Delta r_i$ are the radial mesh and grid spacing
of the time-dependent hydrodynamic code, $F_{i-1/2}$ and
$h_{i-1/2}$ are the radiative flux and disc height
(evaluated at the mid-point) and $N$ is the number of radial points
within the ring. Clearly, since the ring luminosity is calculated
as the sum of the contributions from the original zones, it
differs slightly from one annulus to another. In terms of the
width $\Delta r = r_N - r_1$ and the average radius $r_m = (r_1 +
r_N)/2$ of each ring, the luminosity can be approximately written
as

\begin{equation}
L_r = 2\pi r_m \Delta r r_G^2 \sum_i F_i/N \,. \label{lumring}
\end{equation}

We found that the aforementioned zoning procedure guarantees a
sufficient accuracy when summing up all the rings and comparing the
total disc luminosity to that computed from the output of the transfer
calculation. However, especially in model 2, the temperature may
change significantly within a ring and this contrasts the assumption,
which will be introduced later on, that it may be approximated as
isothermal. Therefore, we further ask that the fractional temperature
variation in each ring does not exceed $2.5\%$, independently of the
zone luminosity. Because the outermost zones of the original grid do
not contribute appreciably to the total luminosity, they are grouped
together into one single broad ring. Typically, this procedure results
in a total number of rings $N_{rings} = 30-40$ (see table \ref{tabmod}
for details).

The calculation of the radiation spectrum from each annulus
proceeds in a way similar to that of Shimura \& Takahara
(\cite{st93:1993}; \cite{st95:1995}) for Shakura-Sunyaev discs.
Each ring is characterized by its luminosity and height above the
mid-plane, and is approximated as (radially) isothermal. No
energy dissipation occurs in the disc atmosphere that is then in
radiative energy equilibrium. All the radiative flux is produced
in the disc and enters the atmosphere from below. Consequently,
the radiative luminosity is constant throughout the atmosphere,
$L = L_r$.

The vertical structure is computed solving the coupled
hydrostatic, energy and transfer equations in a plane-parallel,
completely ionized, hydrogen layer. Solving the transfer of
radiation in a geometrically thin {\it plane parallel} slab is
formally very similar to solving the corresponding problem in a
geometrically thin {\it radial} atmosphere. Thus we decided to
model the disc atmosphere using the same equations considered by
Zampieri et al. \cite{ztzt:1995} for a static, spherically
symmetric, hydrogen atmosphere around a neutron star. The
equations have been slightly modified neglecting general
relativistic corrections, taking into account for the radial
displacement of the ring from the center of gravity and suppressing the
energy release within the atmosphere. The
scattering depth $\tau = -\int_{h_r}^\infty \kappa_{es}\rho\,dz$
(here $\kappa_{es}$ is the scattering opacity and $\rho$ the
density) is used as the independent variable and the inner and
outer boundaries are placed at $\tau_{in}=8$ and
$\tau_{out}=10^{-4}$, respectively. The variation of the height
$z(\tau)$ is obtained by differentiation. The radiation spectrum
is computed from the first two moment equations for the
monochromatic radiation energy density $U_\nu$ and flux $F_\nu$,
assuming that at the base of the atmosphere, where the $\tau_{eff}
> 1$, the radiation field is Planckian. This set of
coupled differential equations provides the run of pressure
$P$, temperature $T$, height $z$, radiation energy density
$U_\nu$ and flux $F_\nu$ as functions of depth. The density is
obtained from the perfect gas equation of state. A thorough
discussion of the numerical method adopted to solve
these equations can be found in Nobili, Turolla \& Zampieri
\cite{ntz:1993}.

We note that the luminosity computed by the transfer code is $L_{r,sph}=
4\pi
r_{sph}^2 r_G^2 \int F_\nu d\nu$, where $r_{sph}$ is an ``effective
spherical radius''. Assuming $\int F_\nu d\nu = \sum_i F_i/N$
and $L_{r,sph}=L_r$ (equation [\ref{lumring}]), we obtain

\begin{equation}
r_{sph} = \sqrt{r_m \Delta r /2}\, .
\end{equation}
The spectrum emitted by the whole disc is evaluated by summing up
the contributions from the various rings

\begin{equation}
L_{\nu,disc} = \sum_{rings} 4\pi r_{sph}^2 r_G^2 F_\nu\, .
\end{equation}

\begin{table}
\caption{Model parameters.\label{tabmod}}
\begin{tabular}{@{}ccccc@{}}
   Model   &  $L_{disc}/L_E$   & Radial zones & $f_{ave}$ &   \\[10pt]
     1   & $4.8\times 10^{-2}$ &  33 & 1.64 &  \\
     2   &  1.54               &  41 & 1.65 &  \\
     3   & $1.6\times 10^{-2}$ &  29 & 1.65 &  \\
     4   & $2.4\times 10^{-2}$ &  32 & 1.65 &
\end{tabular}
\end{table}

\section{Spectra from transonic accretion discs}
\label{result}

\begin{figure*}
\centering \vspace{12.cm} \includegraphics{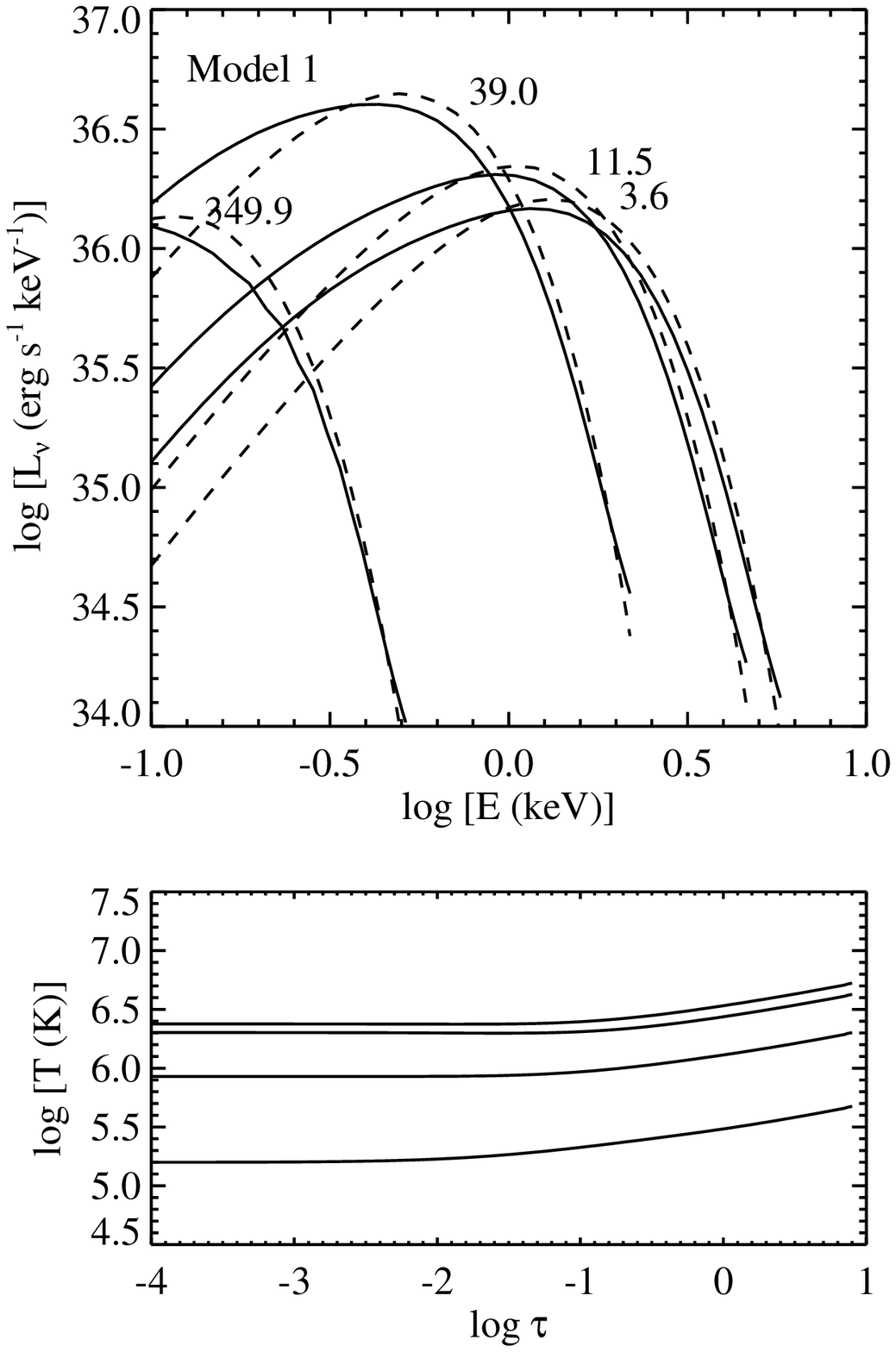}
\includegraphics{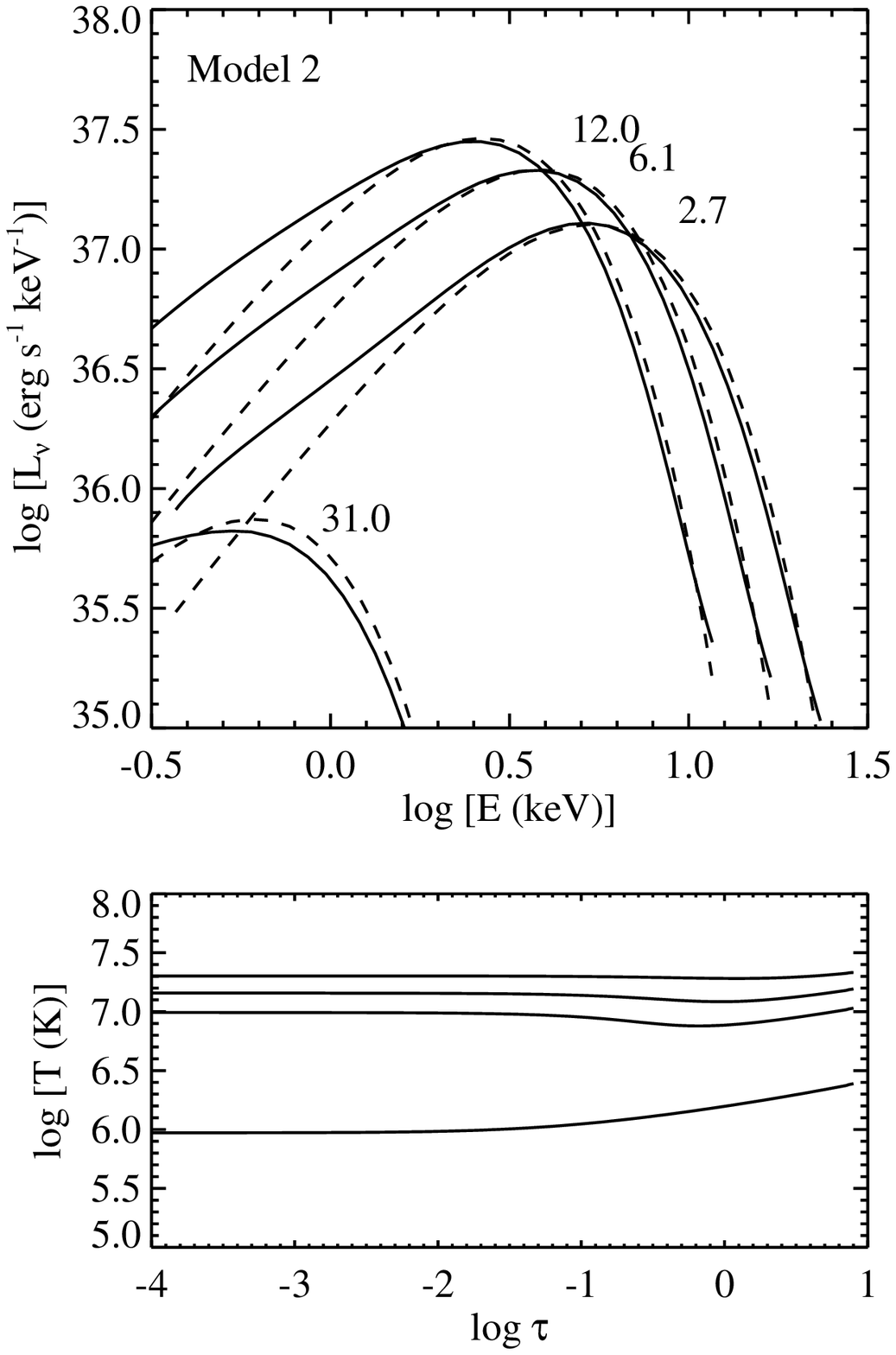} \caption{\label{local12}{Upper
panel: local spectra at different radii (full lines) together
with the best-fitting blackbody (dashed lines) for model 1 (left)
and model 2 (right); lower panel: the run of temperature in the
atmosphere vs. scattering depth. }}
\end{figure*}

\begin{figure*}
\centering\vspace{12.cm}\includegraphics{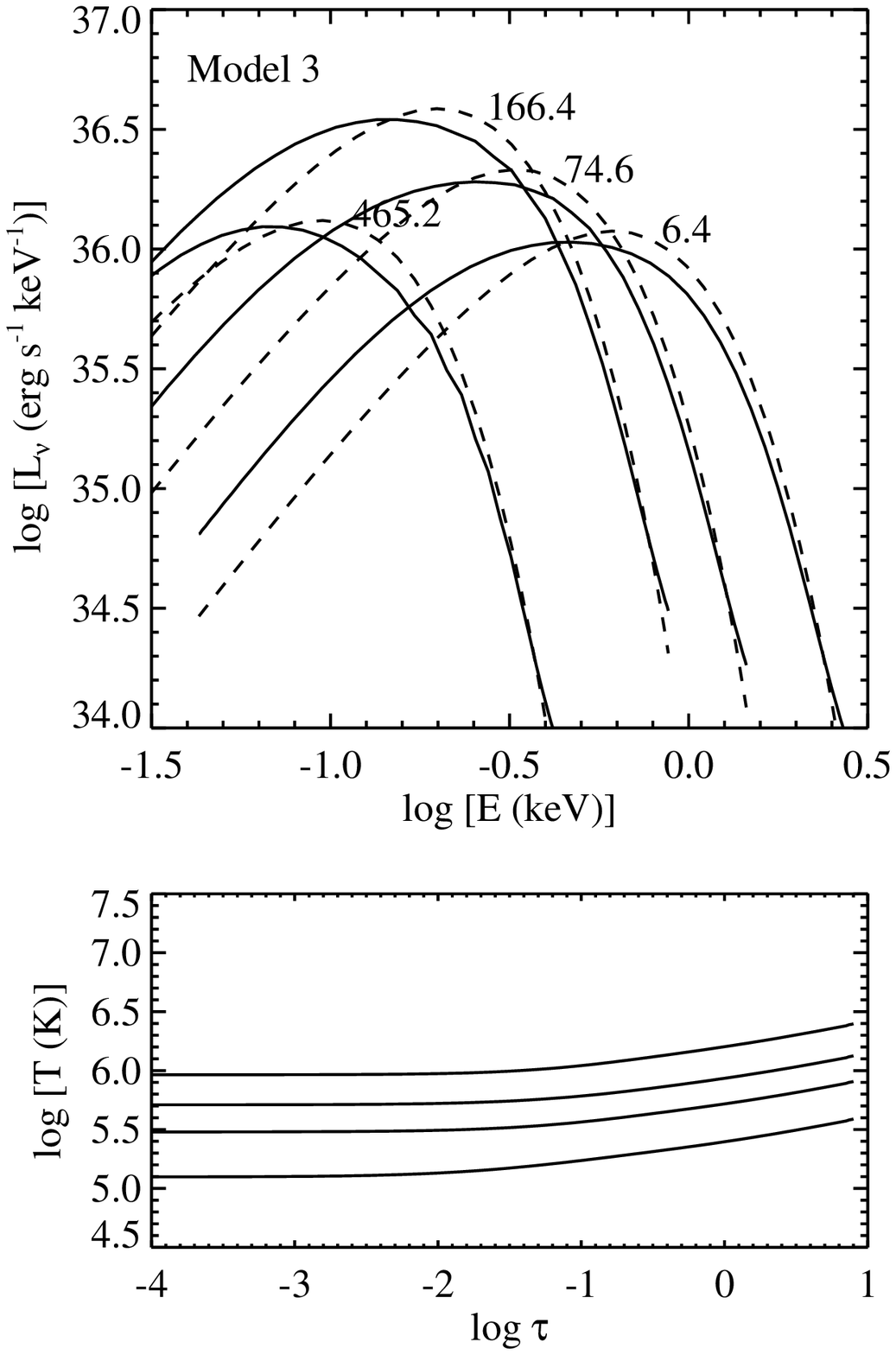}
\includegraphics{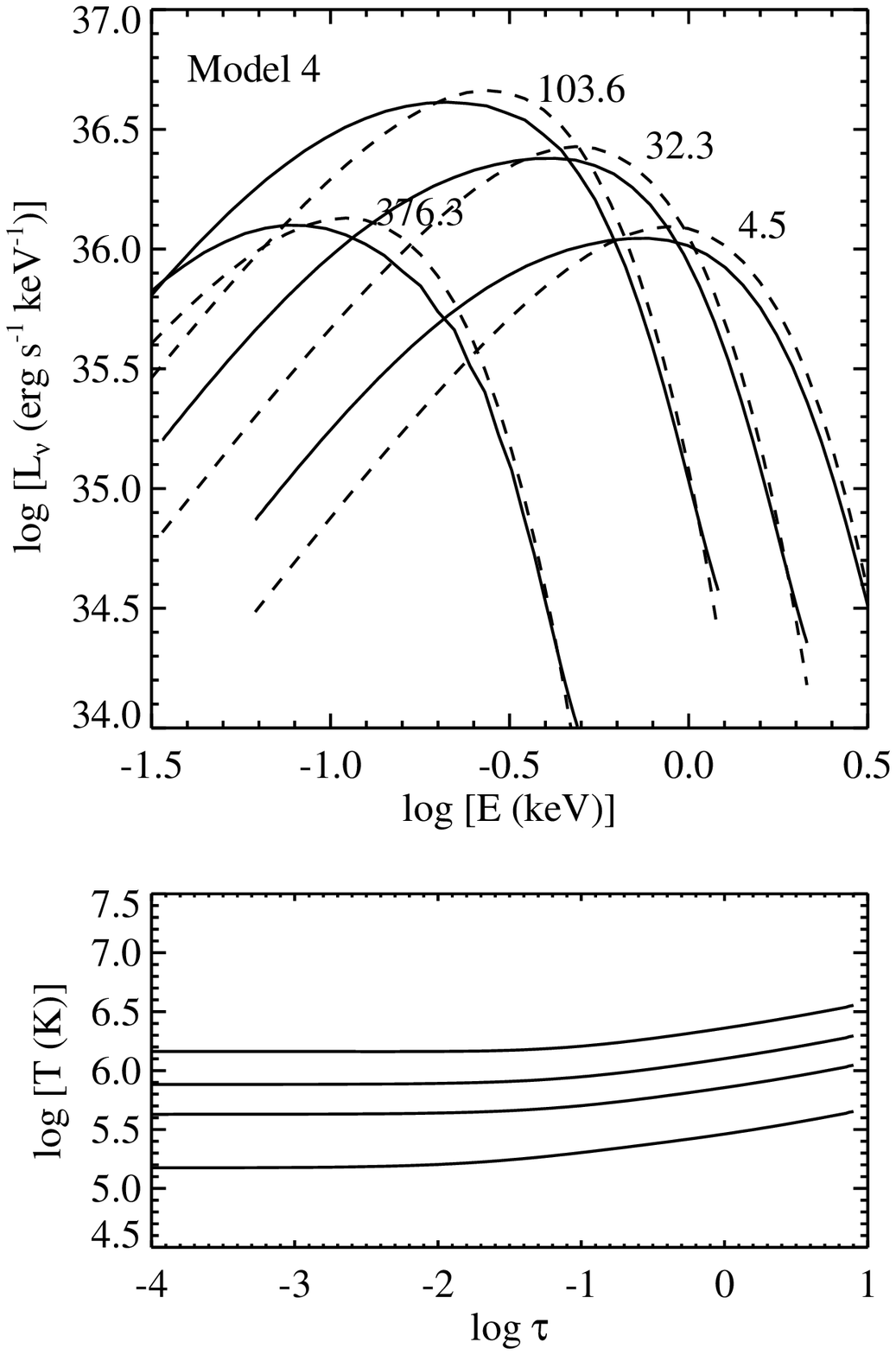} \caption{\label{local34}{Same as in
Fig. \ref{local12} for model 3 (left) and model 4 (right).
        }}
\end{figure*}

Local spectra from a number of selected radial zones are shown in
Figs. \ref{local12} and \ref{local34} for the four evolutionary
stages considered, together with the corresponding temperature
profiles in the disc atmosphere. We found that the emergent
spectra are harder than the blackbody at the ring effective
temperature, $T_{eff,r} = (L_r/ 2\pi r_m r_G^2\Delta
r\sigma)^{1/4}$ (here $\sigma$ is the Stefan-Boltzmann constant).
The presence of an overall spectral hardening is a common feature
of radiative transfer in plane-parallel atmospheres and has been
discussed by various authors in different contexts (e.g. London,
Howard \& Taam \cite{lht:1986} for atmospheres around X-ray
bursting neutron stars; Romani \cite{ro:1987} and Zampieri et al.
\cite{ztzt:1995} for atmospheres around cooling and accreting
neutron stars; Shimura \& Takahara \cite{st95:1995} for the
reprocessing of radiation by the top layers of Shakura-Sunyaev
discs).

The hardening is essentially caused by two distinct effects.
Models 1, 3 and 4 have an average ${\dot M} \sim 0.5$ (in Eddington units)
and are rather cool (atmospheric temperature $\la 5\times 10^6$ K). Their 
deeper layers
($\tau\sim \tau_{in}$) are
comparatively hotter, $T(\tau_{in}) \ga T_{eff}$. High energy photons
produced there escape more easily, because the free-free opacity
(which is the main source of absorption) declines rapidly with
increasing frequency. The emergent spectrum is then, roughly, the
superposition of blackbody spectra with decreasing temperatures, which
is peaked at $\approx kT(\tau_{in})$. This mechanism is responsible
for the spectral broadening with respect to a Planckian and the
sensible deviation from the best-fitting blackbody in the local
spectra of models 1, 3 and 4 (see Fig. \ref{local34} and the upper
left panel in Fig. \ref{local12}).

The case is different for model 2, which is more luminous and
comparatively hotter, $T(\tau_{in})\la 3\times 10^7$ K, with an
average accretion rate ${\dot M} \sim 25$ (in Eddington units).
Local spectra at small radii
are more affected by scattering and they appear nearly Planckian close
to the maximum. The scattering depth of the inner layers
($\tau_{in}\sim 10$) is large enough to build up a Compton parameter
$y=4(kT/mc^2)\max(\tau, \tau^2)\sim 0.1-1$. Therefore, high energy
photons tend to fill the Wien peak at the local atmospheric
temperature and photons with energy $\ga 3kT$ are systematically
downscattered. As Shimura \& Takahara
\cite{st95:1995} pointed out, this explains the nearly Planckian
shape of the spectrum close to the maximum and also the low
energy tail in excess to the best-fitting blackbody clearly
visible in the first three spectra in Fig. \ref{local12} for
model 2. The Compton parameter of models 1, 3 and 4 is a factor
$\sim 10$ smaller, $y\la 0.1$, not large enough to produce any
spectral distortion.

\begin{figure}
\vskip 6.2cm
\includegraphics{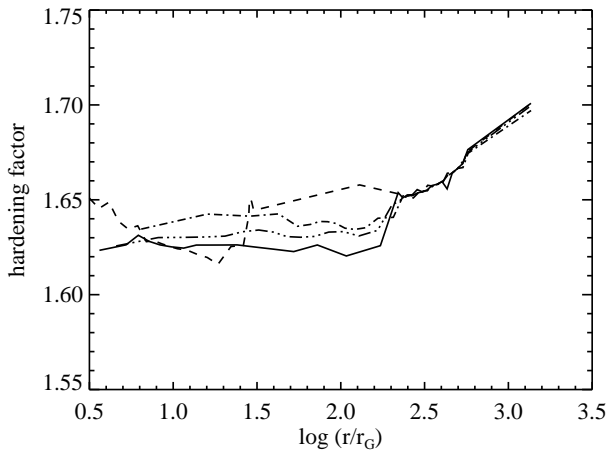}
\caption{\label{hard}{The radial variation of the
hardening factor for models 1-4 (full, dash, dash-dotted, dash-double dotted
line respectively).
}}
\end{figure}

\begin{figure*}
\centering
\begin{minipage}{140mm}
\vspace{340pt} \includegraphics{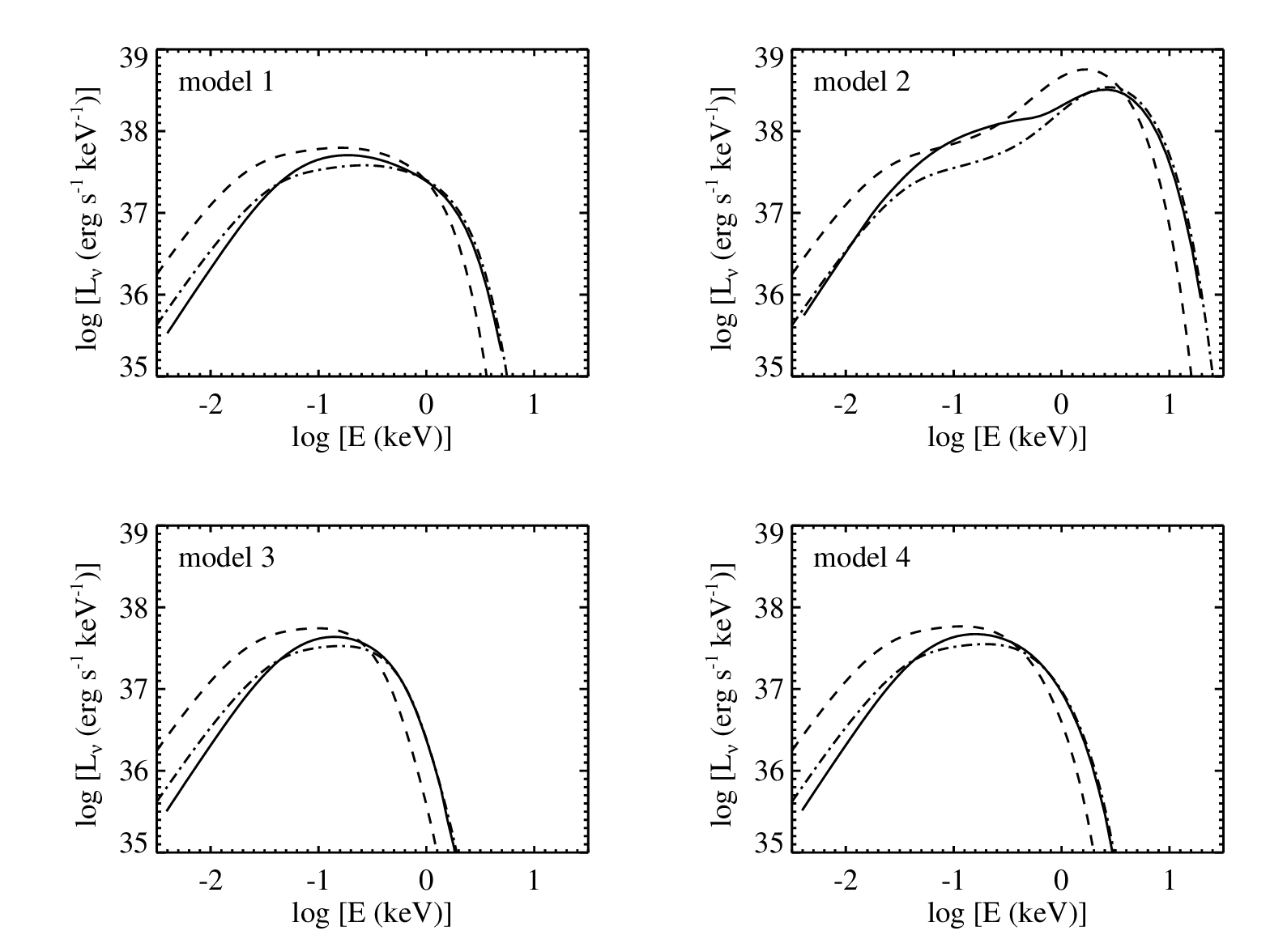}
\caption{\label{spectot}{Spectra emitted by the whole disc (full
lines)
 together with the multicolour blackbody spectra at $T_{eff}$ (dashed lines)
        and at $f_{ave}T_{eff}$ (dash-dotted lines).}}
\end{minipage}
\end{figure*}

Despite the fact that local spectra may deviate appreciably from a
blackbody, it is quite useful to introduce, as is commonly done, a
hardening factor to quantify the spectral changes caused by
radiative transfer in the disc atmosphere. For each radial zone
the fit of the computed spectrum with a Planckian, $\propto
\nu^3/(\exp{h\nu/kT_c} -1)$, gives the color temperature $T_c$.
The hardening factor is defined as $f = T_c/T_{eff}$ and measures
the overall energy shift of the spectrum with respect to the
blackbody at $T_{eff}$, regardless of the actual spectral
distortions.  The run of the hardening factor in the disc is
shown in Fig. \ref{hard}.
Quite remarkably, we find that the hardening factor is
very similar in all four evolutionary stages considered
here, and is very nearly constant through the disc. The typical
value is $f\sim 1.65$  and it shows
only a marginal increase in the outer zones. Following Shimura
\& Takahara \cite{st95:1995}, we define a radially-averaged
hardening factor as
\begin{equation}
f_{ave} = \sum_{rings} f_r/N_{rings} \,.
\end{equation}
The values of $f_{ave}$ for our models are listed in table
\ref{tabmod} and the spectra from the whole disc are
shown in Fig. \ref{spectot}, together with the multicolor blackbody
spectra at $fT_{eff}$ with both $f = f_{ave}$ and $f=1$.
As shown in Fig. \ref{spectot}, with the possible exception of the
intermediate energy range of model 2, the spectrum from the whole disc
can be qualitatively described in terms of a superposition of
Planckian spectra with an average temperature calculated using
$f_{ave}$. We note that a similar result was found by Shimura \&
Takahara \cite{st95:1995} for Shakura-Sunyaev discs with intermediate
values of the accretion rate and $\alpha = 0.1$.

The behavior of the hydrodynamic variables across the disc atmosphere
is briefly illustrated here. Because the luminosity of each radial
zone is well below the Eddington limit, pressure scales linearly with
depth. The temperature variation across the atmosphere is rather
modest: $T$ decreases monotonically outwards,
with the exception of the inner zones of model 2, where the
increase is caused by the effectiveness of Compton heating/cooling
in the energy balance (see the bottom panels in Figs.
\ref{local12}-\ref{local34}). The atmospheric scale height is
always quite small and the fractional increase of $z(\tau_{out})$ over
the average ring height $h_r$ is only a few percent.

\begin{figure*}
\centering
\begin{minipage}{140mm}
\vspace{340pt} \includegraphics{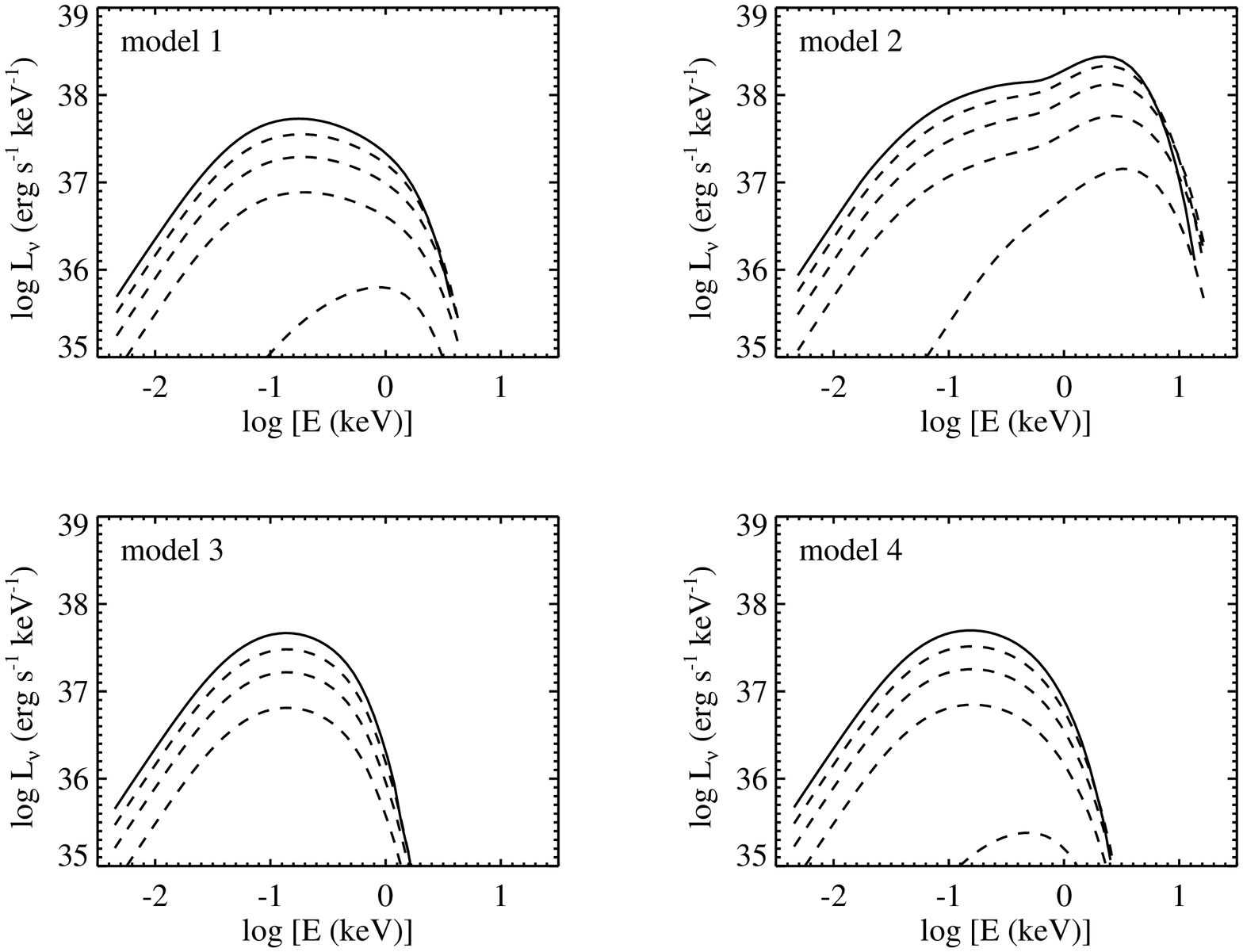}
\caption{\label{specinfty}{Observed spectra at radial infinity
for different viewing angles. The full lines refer to $\cos \theta
= 1$ (disc face-on), whereas the dashed lines show spectra
observed at $\cos \theta = 0.75, 0.5, 0.25, 0$ (from top to
bottom).}}
\end{minipage}
\end{figure*}

Although present spectra can not be taken as representative of
the whole X-ray emission from BHCs, in that they do not exhibit
any power-law hard tail (see section \ref{conclu}), actual
spectra as seen by an observer at infinity have been calculated.
General relativistic effects have been included,
following the method by Cunningham \cite{cun:1975}. We consider
here only the case of a Schwarzschild black hole, to which the pseudo-Newtonian potential
is the closest non-relativistic analog. The same calculation could
be easily performed for a maximally rotating Kerr black hole but this
seems not appropriate in the present framework. The original
assumption of Keplerian rotation has been maintained. Deviations from
a pure Keplerian law in our disc models are not too large and should
not introduce major differences. The specific intensity at the top of
the atmosphere ($\tau_{out}=10^{-4}$), needed to evaluate Cunningham's
expression for the luminosity at infinity, has been obtained in terms
of the first two moments of the radiation field as
\begin{equation}
\label{intensity}
4\pi I_\nu(\mu)= cU_\nu + 3\mu F_\nu = (2+3\mu )F_\nu
\end{equation}
where $\nu$ is the frequency measured by a local observer
comoving with the disc and $\mu =\cos \theta$. Results are shown in Fig.
\ref{specinfty} for different viewing angles, $\theta$. Apart the
obvious decrease in amplitude produced by inclination and
gravitational effects, two major features are apparent in the
spectra at infinity, particularly in the high energy tail which
is produced deeper into the potential well. First, the
exponential cut-off is shifted to lower energies with respect to
the Newtonian model because of the gravitational redshift and,
second, the spectrum tends to become harder for increasing
inclination angle. This is because the orbital motion acquires a
larger component along the line-of-sight and this produces a
larger kinematical blue-shift. Finally, we note that owing to
relativistic ray-bending, radiation from the inner zone can
be seen also when the disc is edge-on ($\cos\theta = 0$).

\section{Discussion and  conclusions}
\label{conclu}

In this paper we have investigated the emission properties of
time-dependent, slim accretion discs, calculating for the first time
the evolution of the disc spectra along the limit instability cycle.
This work is part of an ongoing numerical programme to investigate the
time-dependent behaviour triggered by the onset of various types of
instability (thermal, viscous and acoustic), as outlined in
Szuszkiewicz \& Miller \cite{sm97:1997}, and to get detailed
theoretical predictions for observational quantities relevant to
accreting black holes in galactic X-ray binaries and AGNs, as outlined
in Szuszkiewicz, Turolla \& Zampieri \cite{stz:2000}.

Away from the outburst (maximal evacuation) phase, the spectrum
emitted by each annulus of the disc deviates significantly from a pure
Planckian distribution, showing a characteristic broadening.  This
effect is typical of the spectra emitted by free-free dominated
atmospheres and is caused by the superposition of photons emitted at
different depths: because of the energy dependence of the free-free
opacity, more energetic photons are emitted in the deeper atmospheric
layers, where the temperature is higher. Outburst
spectra are more Planckian in shape around the maximum because the
higher atmospheric temperature makes comptonization comparatively more
effective: photons tend to form an excess at low energies and
to fill the Wien peak at high energies.
A usual way to estimate the overall hardening of a thermal
spectrum is to compare the best-fitting Planckian distribution with a
blackbody at the effective temperature. We found that, in all stages
of the instability cycle, the temperature of the best-fitting
blackbody is typically larger than the local effective temperature by
a factor $f\sim 1.65$, indicating that the hardening factor is
substantially constant both in radius and time. There is only a slight
increase of $f$ with radius in the outer region. Therefore, although
the mechanism responsible for the hardening is different (thermal
comptonization in one case and superposition of blackbody spectra in
the others), outburst spectra have an average hardening factor
comparable to that of the spectra emitted during the evacuation and
refilling stages. We note that the average value of the hardening
factor for the local spectra in the initial state and the
tendency to increase outwards are consistent with what found for a
Shakura-Sunyaev disc by Shimura \& Takahara (\cite{st95:1995}; $f\sim
1.7$) for similar values of the black hole mass, viscosity parameter
and accretion rate. The values of the hardening factor during the
other stages cannot be directly compared because the structure of the
disc is rather different.

In this investigation we focused on the properties of radiation
emitted by thermally-unstable accretion discs. We remind that in the
present model no energy is released in the disc atmosphere and that no
high-energy, thermal or non-thermal, electron component has been
included. Therefore, computed spectra are thermal and do not show the
hard tail often observed in BHCs and AGNs. This must be taken into
account in comparing present models with the observed spectra, in that
only the evolution of the soft component can be reproduced.  As shown
in Figs. \ref{spectot} and ~\ref{specinfty}, our model undergoes
rather dramatic spectral changes along the instability cycle. The
spectrum emitted before the instability sets in is typical of a
standard accretion disc at moderate accretion rates and has a broad
bump in the 0.1--1 keV band.  During the outburst phase, the inner
region becomes hotter and contributes most of the high energy emission
in the 1--10 keV range: a very pronounced peak, centered around $\sim$
5 keV appears in the spectrum. The spectrum then softens considerably
during the refilling phase, up to the point that almost no photons are
emitted above $\sim$ 1 keV. A usual way to describe the observed
spectra, frequently used in X-ray astronomy at low energy resolution,
is to compute the hardness ratios. Because present spectra lack the
power-law, high energy tail, we do not attempt to make a quantitative
comparison with the observed hardness ratios but simply track their
evolution during the cycle. In Table~\ref{tab1} we report the hardness
ratio HR1 (HR2), defined as the ratio of the count rate in the 5-13
(13-60) keV energy band over the count rate in the 2-5 keV band, for
the computed spectra of models 1 and 2 (the values for model 3 and 4
are zero). The calculation has been performed using the RXTE response
function and accounting for interstellar absorption (with a column
density $N_H = 2\times 10^{22}$ cm$^{-2}$). As shown in
Table~\ref{tab1}, the hardness ratios at the beginning of the cycle
are low. However, during outburst they increase reaching interesting
values (HR1 $\sim 1$, HR2 $\sim 0.03$). Table~\ref{tab1} clearly
indicates that the evolution of our model along the instability cycle
is qualitatively very similar to the evolution of the soft component
in GRS1915+105 (see Belloni et al. \cite{bello00:2000}). In this
respect, it is tempting to interpret the observed soft state B
(characterized by a higher temperature) as corresponding to the
outburst spectral state of our models. The higher temperature is
associated to the innermost part of the disc, accreting at much higher
rates than the outer unperturbed part. Further investigation of this
promising result is definitely required and is presently under way.

\begin{table}
 \centering
 \begin{minipage}{80mm}
 \caption{Hardness ratios for the computed spectra of models 1 and 2.}
 \label{tab1}
 \begin{tabular}{ccccc}
                 & \multicolumn{2}{c}{model 1}
                 & \multicolumn{2}{c}{model 2}                \\
  $\cos\theta$   & HR1\footnote{HR1 = (5-10 keV)/(2-5 keV).}
                 & HR2\footnote{HR2 = (13-60 keV)/(2-5 keV).}
                 & HR1$^a$
                 & HR2$^b$                       \\
                 &        &         &              &           \\
  1              & 0.021  & 0       & 0.63         & 0.003     \\
  0.75           & 0.042  & 0       & 0.83         & 0.014     \\
  0.5            & 0.055  & 0       & 1.02         & 0.031     \\
  0.25           & 0.064  & 0       & 1.18         & 0.052     \\
  0              & 0.05   & 0       & 1.31         & 0.054     \\
 \end{tabular}
 \end{minipage}
\end{table}

A few comments on the assumptions involved in our treatment of the
accretion flow and radiative transfer should be made. We decided to
use the slim disc approximation even if we are aware that realistic
discs are truly 2D structures in which all thermodynamic
quantities vary with height and radius. Modeling the
2D structure of accretion discs is much more complicated than
using a 1D vertically averaged approximation. However, as
recently discussed by Hur\'e \& Galliano \cite{hure01:2001}, the
accuracy of vertically averaged models should be
sufficient in practice for many astrophysical applications. The slim
disc model (Abramowicz et al.
\cite{acls:1988}) retains the assumption that the disc is
geometrically thin but the centrifugal force no longer balances
gravity exactly. The thickness of the disc depends on both pressure
and viscosity: the higher is the pressure and (in general) the
viscosity, the thicker is the disc. In going from extremely thin discs
to thicker ones, the effects of advection become progressively more
important. The slim disc approach includes the effects of advection in
a correct way while continuing to make use of vertical
averaging. Advection is particularly important for the inner parts of
an accretion disc around a black hole: while a standard Keplerian disc
ends at the marginally stable orbit, a realistic accretion flow
may continue
 inwards and must cross a sonic point before reaching the
event horizon. Finally, we note that gravity has been represented by
the pseudo-Newtonian potential of Paczy\'nski \& Wiita
\cite{pacz:1980}, which describes in a satisfactory way the dynamical
effects of the Schwarzschild gravitational potential.  The main
approximation that has been introduced in treating the vertical
transfer of radiation in the disc concerns stationarity. This
approximation turns out to be very good almost everywhere in the disc
because the time-scale over which the radiation field changes is much 
shorter than the fastest dynamical time $r/v_{\phi}$,
$t_{rad}=(c_s/c\tau_{eff})  (r/v_{\phi})\ll r/v_{\phi} \ll r/v_r$,
where $c_s$, $v_{\phi}$ and $v_r$ are the sound, azimuthal and
radial velocities.
Only in the transonic
region of the disc this approximation might break down. However, the
radial extent of this region is usually fairly small and does not
contribute significantly to the total disc luminosity ($\la 5\%$ for
the outburst state).  The disc model used here neglects all the
effects induced on the surface layers by the X-ray photons produced
(or scattered) by an external medium. X-ray illumination by an
external source heats the disc photosphere and leads to the formation
of a highly ionized layer which can become Thomson thick for large
enough luminosities (see e.g. Nayakshin, Kazanas \& Kallman
\cite{nkk00:2000}; Nayakshin \& Kallman \cite{nk01:2001}). While this
will definitely influence the emitted spectrum shifting the emission 
at higher energies, its effect onto the hardening factor are more
difficult to estimate and would require a consistent calculation.
We only note that present results show that $f$ is not much sensitive to 
the disc surface temperature (see Table \ref{tabmod}).

\section*{Acknowledgments} E.S. gratefully acknowledges financial
support from the Polish State Committee for Scientific Research (grant
KBN 2P03D01817), the Italian Ministry for University and Scientific
Research (MURST) and the University of Padova. Work partially
supported by MURST through grant COFIN-98-02154100.

\label{lastpage}

\end{document}